\documentclass[preprint2]{aastex}

\shorttitle{A highly-collimated SiO jet in the HH212 protostellar outflow}
\shortauthors{Codella et al.}

\begin{document}


\title{A highly-collimated SiO jet in the HH212 protostellar outflow}


\author{C. Codella}
\affil{INAF - Istituto di Radioastronomia, Sezione di Firenze, Largo E. Fermi 5, 50137
Firenze, Italy}
\email{codella@arcetri.astro.it}

\author{S. Cabrit}
\affil{LERMA, UMR 8112 du CNRS, Observatoire de Paris, 61 Av. de l'Observatoire, 75014 Paris,
France}
\email{sylvie.cabrit@obspm.fr}

\author{F. Gueth}
\affil{IRAM, 300 rue de la Piscine, 38406 Saint Martin d'H\'eres, France}
\email{gueth@iram.fr}

\author{R. Cesaroni and F. Bacciotti}
\affil{INAF-Osservatorio Astrofisico di Arcetri, Largo E. Fermi 5,
50125 Firenze, Italy}
\email{cesa@arcetri.astro.it; fran@arcetri.astro.it}

\author{B. Lefloch}
\affil{Laboratoire d'Astrophysique de l'Observatoire de Grenoble, BP 53, 38041 Grenoble Cedex,
France}
\email{lefloch@obs.ujf-grenoble.fr}

\and

\author{M.~J. McCaughrean}
\affil{University of Exeter, School of Physics, Stocker Road, Exeter EX4 4QL, UK}




\begin{abstract}
{We mapped the inner 40$^{''}$ of the HH212 Class 0 outflow in
SiO(2--1), SiO(5--4) and continuum using the Plateau de Bure interferometer
in its extended configurations. The unprecedented
angular resolution (down to 0$\farcs$34) allows accurate comparison with
a new, deep H$_2$ image obtained at the VLT.
The SiO emission is confined to a highly-collimated bipolar jet
(width $\sim$ 0$\farcs$35) along the outflow axis.
The jet can be traced down to within 500 AU of the protostar,
in a region that is heavily obscured in H$_2$ images. Where both
species are detected, SiO shows the same overall kinematics and
structure as H$_2$, indicating that both molecules are tracing the
same material. 
We find that the high-velocity SiO gas is not tracing a wide-angle
wind but is already confined to a flow inside a narrow cone of
half-opening angle $<$6$\degr$ at $\leq$ 500~AU from the protostar.
Transverse cuts reveal no velocity gradient compatible
with jet rotation above 1 km s$^{-1}$, in contrast to previous claims based
on H$_2$ spectra.}

\end{abstract}


\keywords{Stars: formation -- Radio lines: ISM -- ISM: jets and outflows --
ISM: molecules -- ISM: individual objects: HH212}



\section{Introduction}

The launching of jets from young stars is one of the most enigmatic and
intriguing phenomena in astrophysics. One fundamental problem to
which jets are believed to provide a solution is the removal
of angular momentum from the central accreting system, which is
required to permit low angular momentum material to fall onto the
star.  Recent optical observations \citep[e.g.][]{bacciotti}
suggest that atomic jets from
evolved T-Tauri stars may indeed transport angular momentum away
from the inner regions of the accretion disk.
Is the same mechanism at the origin of
the molecular jet counterparts observed when the
protostar is still deeply embedded in its natal high-density core?
To address this question, we have investigate HH212 in Orion ($d$ = 460 pc),
a highly
symmetric bipolar H$_2$ jet 
\citep{zinne98}, associated with a collimated CO outflow 
\citep{lee}, and driven by a low-luminosity Class 0 protostar, IRAS05413--0104.
HH212 provides an optimal situation in which to study
jet kinematics, since it
lies close to the plane of the sky \citep[4$\degr$;][]{claussen}, 
shows hints of rotation in one H$_2$ knot 
\citep{davis}, and is surrounded by a compact ammonia core
rotating about an axis aligned with the jet 
\citep{wiseman}. As a tracer, we used SiO, which is associated with
shocks and generally suffers
minimal contamination from infalling envelopes or swept-up cavities
\citep{guilloteau,hirano}.

\section{Observations}

The SiO observations of HH212 were obtained in January-March 2006
with the IRAM Plateau de Bure Interferometer (PdBI) in France.  Two
6-element configurations were used: the new extended A configuration,
which includes baselines from 136~m up to 760~m, and the new B
configuration, with baselines from 88~m to 452~m. The dual-channel
receivers and the correlator were tuned to simultaneously observe the SiO
$J$ = 2--1 line at 86.847 GHz (resolution: $\sim$ 0.13 km s$^{-1}$;
bandwidth: $\sim$ 70 km s$^{-1}$) and the $J$ = 5--4 line at 217.105 GHz
(0.11 and 55 km s$^{-1}$).  Continuum emission at both frequencies was also
measured.  Amplitude and phase were calibrated by observing 0528+134
and 0605$-$085. The weather conditions were excellent (phase noise rms
of 30$^\circ$ at 86~GHz on the 760~m baselines), leading to a radio
seeing of 0$\farcs$35. The flux density scale was derived by observing
3C\,84 and 3C\,279, with an uncertainty of $\sim$ 25\%.  Images were
produced using natural weighting, and restored with clean beams of
$1\farcs89\times0\farcs94$ (PA=22$^\circ$) at 3.5 mm, and of
$0\farcs78\times0\farcs34$ (PA=21$^\circ$) at 1.4 mm, using the GILDAS
software. The high degree of elongation of the beam is due to the zero
declination of the source. However, the spatial resolution is highest
perpendicular to the jet direction, making the present observations
ideal for the study of the jet collimation and kinematics.

A new, deeper, and higher spatial resolution NIR image of HH212 was
obtained using ISAAC on the ESO Very Large Telescope
(VLT), Paranal, Chile on November~12, 2005 under good conditions.
Although the knots have proper motions of 100--200 km s$^{-1}$ 
\citep{mccaugh02}, the small epoch difference to the SiO data
ensures that the knots have moved only $\sim$
0$\farcs$025, thus facilitating an
comparison between the two data sets. A series of images were taken
through a 1\% wide filter centred on the v=1--0 S(1) line of H$_2$ at
2.122\,$\mu$m. with a total integration time of 32 minutes per
pixel after mosaicing. The image scale was 0$\farcs$147 pixel$^{-1}$.
and the spatial resolution is 0$\farcs$35 FWHM, similar to our SiO(5--4) map.
Accurate astrometry (0$\farcs$14 rms) was derived via 21 stars in
common with the 2MASS Point Source Catalog. The images have not
been continuum-subtracted and thus stellar sources are visible.
However, the HH212 jet itself consists almost exclusively of shocked
H$_2$ line emission and thus there is no confusion along the jet axis.

\section{Results and Discussion}

Emission maps of SiO $J$ = 2--1 and 5--4 are shown in Fig.~1a-b,
superimposed onto the nearly contemporaneous H$_2$ image.  The SiO
emission is confined to a highly collimated bipolar jet, located along
the main outflow axis. The red- and blue-shifted lobes consist of two
symmetric pairs of bullets: an outer pair (Blue 1 and Red 1)
that follows the H$_2$ intensity distribution beyond 10$\arcsec$ from
the driving source, and a new inner pair within $\pm 2\arcsec$ of the
source (Blue 2 and Red 2), with no H$_2$ counterpart. Figure~1c shows
the inner 5$\arcsec$ of the 1.4 mm continuum map, with the
SiO(5--4) emission superimposed.

\subsection{Continuum sources}

The 1.4 mm continuum map (Fig. 1c) shows a bright source (hereafter
called MM1) at position $\alpha_{\rm J2000}$= 05$^{\rm h}$ 43$^{\rm
m}$ 51$\fs$41, $\delta_{\rm J2000}$= --01$\degr$ 02$\arcmin$
53$\farcs$160, in excellent agreement with the VLA position at 3.5 cm
reported by \citet{galvan}. Comparison with
the SiO(5--4) map clearly indicates that MM1 is the driving source of the
molecular jet. In addition, we have a tentative detection at S/N
$\simeq$ 6 ($\sim$ 4 mJy) of a secondary peak (hereafter MM2) at
$\Delta\alpha$=+1$\farcs$6, $\Delta\delta$=--0$\farcs$5, which could
trace an embedded companion. Only MM1 is bright enough to be detected
at 3.5 mm, with a flux of 6 mJy.  Gaussian fitting in the UV plane
at 1.4 mm shows that MM1 is associated with an unresolved peak
($\le$ 0$\farcs$26)
with a faint extension towards SE.  The integrated flux at
1.4 mm is 33 mJy, significantly lower than the 110 mJy reported by 
\citet{lee} in a 2$\farcs$5 beam, indicating that most of the
envelope emission has been resolved out by the PdBI.
Contamination by free-free emission from the jet is negligible: if
we extrapolate the flux measured by
the VLA at 3.5 cm \citep{galvan} with a typical
$\nu^{0.6}$ slope, the expected
contribution is only 0.16 mJy at 3.5 mm.  Hence we are dominated by
dust continuum. The 1.4 mm flux corresponds to a mean surface
brightness of 3.4~K in our beam.  Since
envelopes and disks are not elongated along the jet axis, the source size is less
than 0$\farcs$26 $\times$ 0$\farcs$26 and thus the true brightness temperature is
quite high: $\ge$ 13~K.
The spectral slope between 1.4
mm and 3.5 mm, derived from UV data, is $\sim$ 2.
Both facts seem to indicate that dust emission is close to optically
thick at 1.4 mm. This would suggest that the unresolved mm peak is
not tracing the inner parts of the envelope, but a circumstellar disk
viewed close to edge-on. The disk diameter would be $\le$ 0$\farcs$26
= 117~AU.

\subsection{The inner SiO bipolar jet}

Our high resolution maps reveal a small-scale pair of SiO emission knots
emerging from MM1 and extending out to 1$\arcsec$--2$\arcsec$ = 500--1000 AU
(Fig.~1c). No H$_2$ counterpart is visible, as the whole structure lies
entirely in the high extinction region around the source. The
beam-deconvolved transversal size of the two lobes (Blue 2/Red 2) is $\sim$
0$\farcs$35, and are thus much more collimated than the hourglass-shaped
bipolar cavity seen in $^{13}$CO on similar scales by 
\citet{lee}.  Position-Velocity (PV) diagrams along the jet axis
(Fig.~2) further show that the SiO emission extends to high
velocities of $\pm$10 km s$^{-1}$ from the ambient velocity \citep[+1.6 km
s$^{-1}$;][]{wiseman}, compared to only $\pm$ 1.7 km
s$^{-1}$ for $^{13}$CO \citep{lee}. Hence the SiO lobes include
a narrower and faster jet-like component distinct from the swept-up
cavity. The maximum radial velocity is similar to the typical centroid
velocity of H$_2$ knots further out (see Fig.~2), arguing that the
high-velocity SiO is probably tracing the base of the large-scale
($\sim$ 0.5 pc) molecular jet.
Correction for inclination 
\citep[4$\degr$ to the plane of the sky;][]{claussen}
gives a full jet speed $\sim$ 140 km
s$^{-1}$, comparable to the proper motion of H$_2$ knots 
\citep{mccaugh02}. Inner SiO knots were also recently imaged in HH211 down to $\le
2\arcsec$ from the protostar \citep{hirano,gueth06}. 
Hence SiO appears to be a powerful tracer of the jet
base in Class 0 sources.

\citet{hirano} found broad SiO linewidths in the protostar
in HH211, which they attributed to a wide angle wind. Such an
interpretation is not supported in HH212, however: SiO profiles show
little blue/red overlap in each lobe (see Fig.~2), despite the almost
edge-on inclination.  Assuming a conical wind of constant speed, a
half-opening angle $\le$ 6\degr\ is inferred, ruling out a wide-angle
wind. The fact that SiO emission stops {\it precisely} near zero
velocity, with an overlap of at most 2 km s$^{-1}$, would actually suggest
that the wind half-opening angle coincides to
within 2\degr\ with the angle from the plane of the sky\footnote{The
wind half-opening angle, $\theta_{\rm max}$, is related to the angle
to the plane of the sky, $\alpha$, by $\tan\theta_{\rm max} =
(R-1)/(R+1) \tan\alpha$ where $R$ is the ratio (in algebraic value) of
maximum to minimum radial velocities in a given lobe. In HH212,
$v_{\rm max} > 10$ km s$^{-1}$, $-2<v_{\rm min} <0$
km s$^{-1}$, and $\alpha$=4\degr\ yield $R<-5$ and $\theta_{\rm max}$ =
4\degr-6\degr.}. Since such a coincidence is statistically unlikely,
it appears more probable that high-velocity SiO is confined to {\it
less} than $4$\degr\ from the jet axis (thus producing no emission at
zero velocity) and that SiO emission near systemic velocity is tracing
{\it intrinsically slower}, possibly less collimated material.
This slow material may trace unresolved bowshock
wings. Indeed, water maser spots observed via VLBI reveal curved
bowshocks at $\pm$ 0$\farcs$1 from the protostar, covering a range of
radial velocities \citep[black arrows in the right panel of Fig. 2;][]{claussen}. 
Given the restrictive excitation and coherence
conditions for maser amplification, the observed velocity range is
less than the full range in the bow. Hence, internal
bowshocks could contribute significantly to the broad SiO line
widths. Intrinsic gradients in jet speed (due e.g. to time
variability, or to a range of launch radii) may also be present.

\subsection{The outer SiO bullets}

At distances of 10$\arcsec$ and beyond, the spatial correspondence
between SiO and H$_2$ is very good. The two symmetric elongated SiO
bullets (Blue 1/Red 1) peak towards H$_2$ knots SK2-SK4 and NK2-NK4
(note that the SiO(5--4) image is affected by primary beam
attenuation at large distances from the centre (FWHM = 22$\arcsec$).
SiO emission is also seen
between and beyond these knots, with a morphology similar to that in
H$_2$ (Fig. 1).  The correspondence between SiO and H$_2$ in the outer
bullets is also excellent in radial velocity, as shown in the PV
diagram in Fig. 2 (left panel).  The SiO(2-1) profiles peak within 2
km s$^{-1}$ of the H$_2$ centroid. Hence SiO appears to trace the same
jet as H$_2$, confirming the conclusion of 
\citet{takami} based on SiO(8--7) profiles at much lower
angular resolution (22$\arcsec$). The SiO line profiles are broader
than towards the inner knots, but they show the same extension down to
zero velocity and small degree of red/blue overlap, suggestive of
internal bowshocks. Line profile asymmetries between the two lobes
indicate slight differences in shock structure despite the highly
symmetric knot spacings.

In contrast to the excellent agreement between SiO and H$_2$ beyond
10$\arcsec$, we find a striking lack of SiO emission towards the
brightest H$_2$ knots (SK1, NK1). A similar effect was observed
towards the HH211 jet, where the brightest, bow-shaped H$_2$ knots
also lack an SiO counterpart \citep{hirano}.  One
possible explanation would be that SK1 and NK1 trace more powerful
shocks where SiO does not form/survive or where the SiO
excitation is extremely high and consequently the low-J transitions are
very weak. The detection of [Fe~II] emission in SK1 and NK1
\citep{zinne98,cog},
their clear bow-shock geometry (see Fig.~1), and
their 3-4 times broader linewidth compared to other H$_2$ knots in
HH212 (Fig.~2) support this hypothesis. Estimates of SiO
excitation and abundances will be presented in a separate forthcoming
paper.

\subsection{Search for jet rotation}

\citet{davis} reported a centroid velocity shift of
+2.3 km s$^{-1}$ from NW to SE across knot SK1 which they attributed
to jet rotation, based on its agreement with the rotation sense of the
ammonia protostellar core \citep{wiseman}. Although
SK1 is not detected in SiO (see 3.3), we can test this interpretation
in the other knots, as rotation speeds should remain roughly
constant once the jet has achieved cylindrical collimation
\citep[see][]{pesenti}.

Figure~3a presents PV diagrams across the jet in SiO(5-4) for each
emitting region. No shift of the intensity peak is seen across Blue 1
on the scales $\pm$ 0$\farcs$5 investigated by 
\citet{davis}. On the other hand, a clear gradient is observed
across the Red 1 bullet (Fig. 3, bottom panel), with a shift of --6 km
s$^{-1}$ over $\sim$ 1$\arcsec$ from NW to SE. However, the gradient
goes in opposite sense to the rotation pattern of the NH$_3$ core,
ruling out that we are observing rotation of a centrifugally-driven MHD jet.
The fact that
the centroid of Red 1 is slightly displaced from the jet axis, i.e.
that outer SiO bullets are slightly misaligned with
respect to the inner SiO jet, suggests that this effect is due
instead to a kinematical asymmetry in the shocked material, possibly
related to the H$_2$ jet wiggling apparent in Fig.~3b. Since
SK1 traces a much wider bowshock than the SK2-SK4 knots in Red 1 (see
Fig.~1), its transverse gradient measured in H$_2$ could easily be
affected by similar asymmetries, and should be interpreted with great
caution. The innermost SiO knots (Blue 2/Red 2) do not show signs of
transverse velocity shifts above 1 km s$^{-1}$, consistent with the
high-velocity SiO jet being practically unresolved
transversally. Higher angular resolution would be needed to constrain
the level of jet rotation.

\acknowledgments

We wish to thank B. Nisini for helpful discussions and suggestions.
This work is supported by the European Community's Marie Curie
Research Training Network JETSET under contract
MRTN-CT-2004-005592. It has benefited from research funding from the
European Community's sixth Framework Programme under RadioNet R113CT
2003 5058187.

\clearpage



\begin{figure*}
\begin{center}
\includegraphics[angle=0,width=\textwidth]{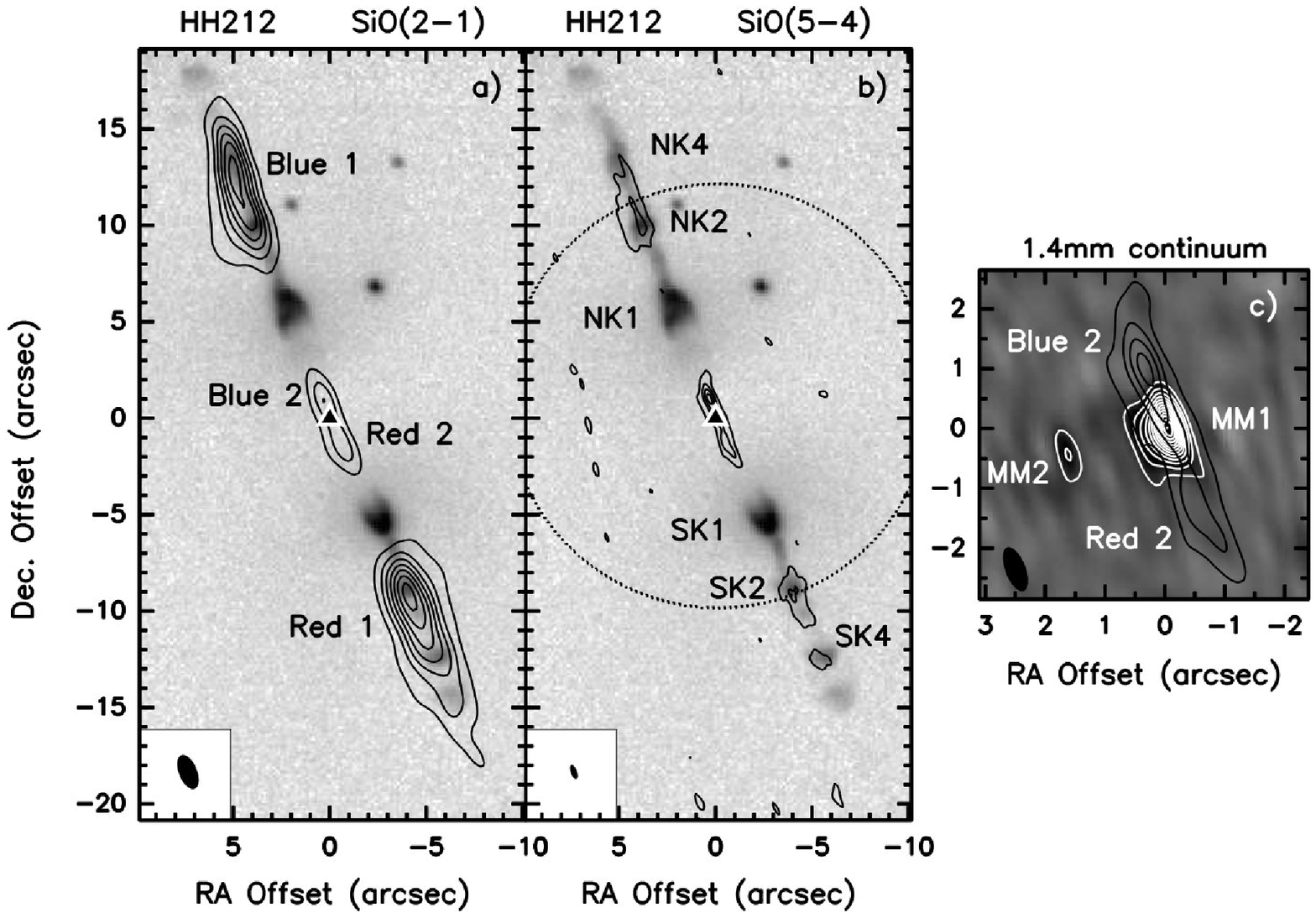}
\caption{{\it a:} Contour map of integrated emission in SiO(2--1)
superimposed on the new H$_2$ image. The SiO emission
regions are labelled in white. The filled triangle at 
(0$\arcsec$,0$\arcsec$) denotes the driving source MM1 
(see Sect.\ 3.1). Contour levels range from
3$\sigma$ (0.22 Jy beam$^{-1}$ km s$^{-1}$) to 39$\sigma$ by steps of 6$\sigma$.
The filled ellipse in the lower left corner shows the
synthesised PdBI beam (HPBW): $1\farcs89\times0\farcs94$.
{\it b:} same as panel ({\it a}) for the SiO(5--4) emission. The dotted circle
indicates the primary beam of the PdBI antennae. H$_2$ knots are
labelled in black. Contour levels range from 3$\sigma$ (0.42 Jy beam$^{-1}$ km s$^{-1}$) to
39$\sigma$
by steps of 3$\sigma$. The beam is $0\farcs78\times0\farcs34$.
{\it c:} Inner 5$\arcsec$ of
the SiO(5--4) map, superimposed onto the continuum image at 1.4 mm (gray
scale and white contours). Continuum contour levels range from
3$\sigma$ (1.9 mJy beam$^{-1}$) to 36$\sigma$ by steps of 6$\sigma$.
Labels indicate the HH212 driving source (MM1) and a
tentative weaker source 
(MM2).[{\it See the electronic edition of the Journal for
a colour version of this figure}.]}
\end{center}
\end{figure*}

\begin{figure*}
\begin{center}
\includegraphics[angle=0,width=\textwidth]{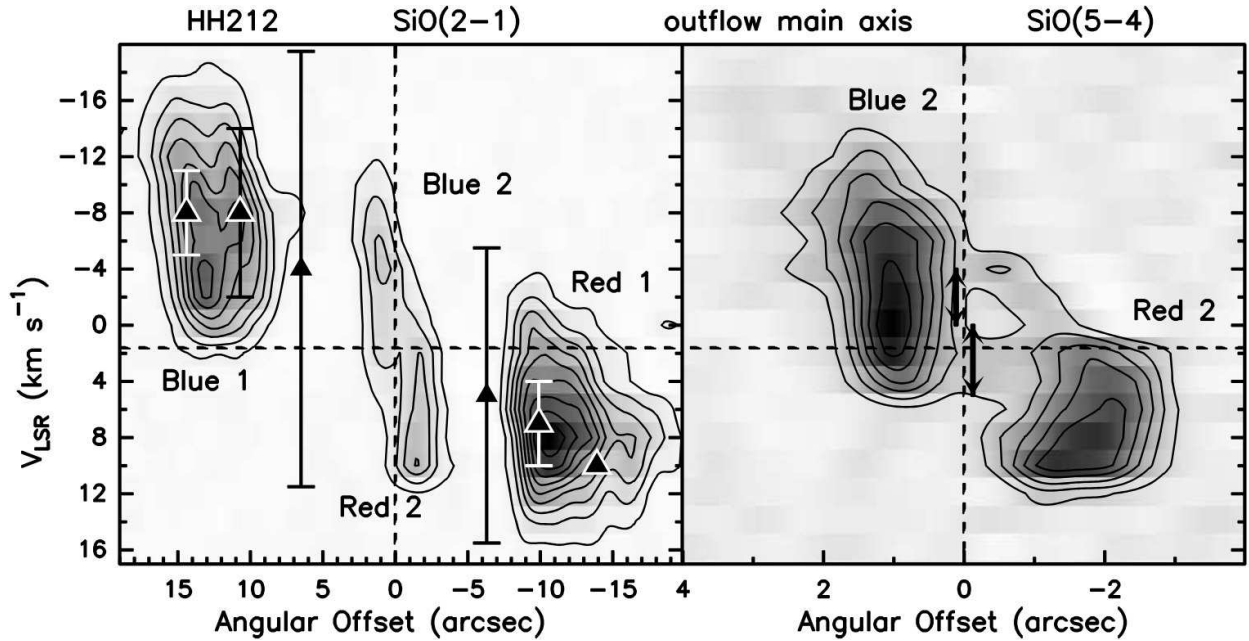}
\caption{{\it Left:} Position-velocity cut of SiO $J$=2--1 along the
jet axis (PA = 22$\degr$). Contour levels range from 1.4~K to 27.5~K by
2.9~K (8$\sigma$). Dashed lines mark the position of MM1 and the cloud
$V_{\rm LSR}$ (+1.6 km s$^{-1}$; Wiseman et al. 2001). Filled
triangles with error bars show the velocity centroid and intrinsic
FWHM of H$_2$ knots (Takami et al. 2006). {\it Right:} Zoomed-in
PV cut of the inner jet in SiO $J$=5--4. Contour levels
range from 3~K to 21~K by 3~K (4$\sigma$). Black arrows show the
velocity range of H$_2$O masers $\pm 0.1\arcsec$ from the
protostar (Claussen et al.\ 1998).}
\end{center}
\end{figure*}

\begin{figure*}
\begin{center}
\includegraphics[angle=0,height=15cm]{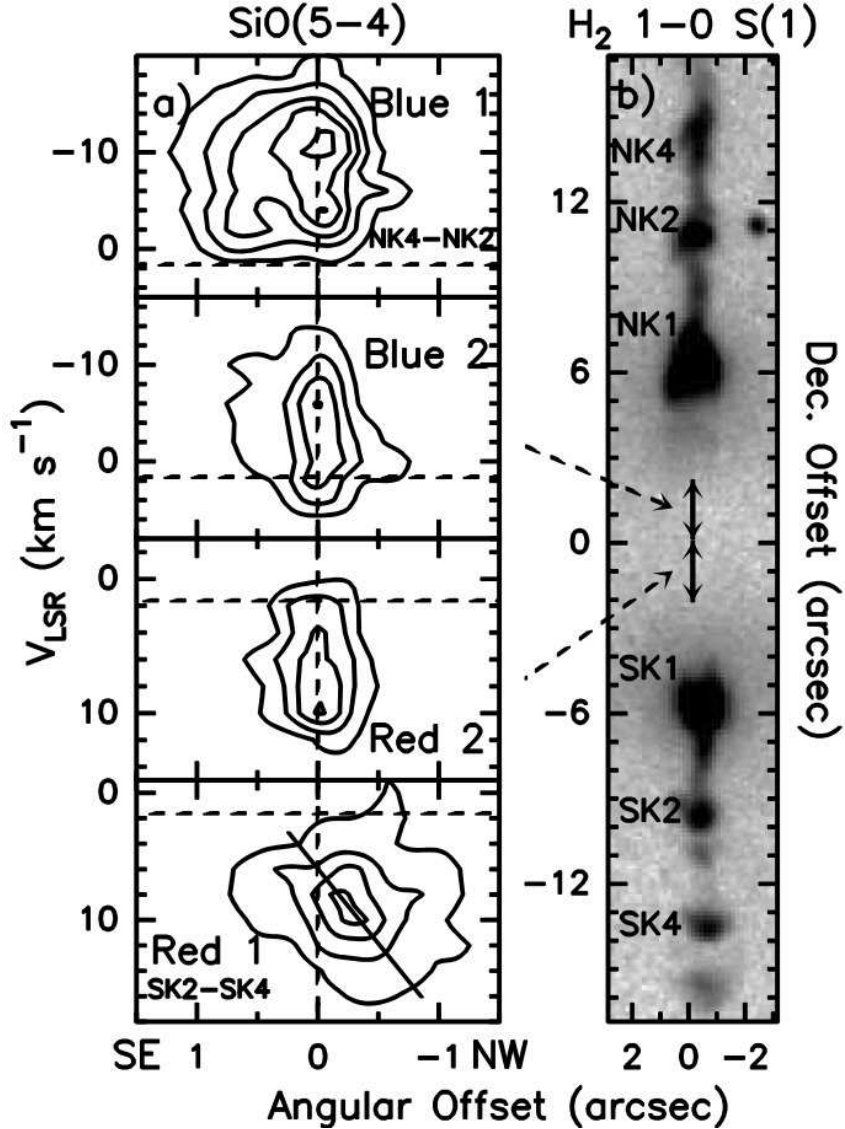}
\caption{{\it a:} PV cuts in SiO(5--4) perpendicular to the jet axis, averaged
over each of the four SiO bullets. Contour levels start at 3$\sigma$
with steps of 6$\sigma$. They range from 1~K to 9.3~K (Blue 1), 2.7
to 14.5~K (Blue 2/Red 2), and 1.7 to 11.6~K (Red 1). Dashed
lines mark the jet axis at PA=22\degr\ and the $V_{\rm LSR}$ of the cloud
(+1.6 km s$^{-1}$; Wiseman et al. 2001). A solid line shows the
velocity gradient across Red~1.
{\it b:} Zoom-in of the H$_2$ image highlighting the jet wiggling.
Vertical black arrows
indicate the declination range of the Blue 2/Red 2 inner jet.}
\end{center}
\end{figure*}

\end{document}